# Effect of Pitch on the Asymmetry in Global Loudness Between Rising- and Falling-Intensity Sounds

Sabine Meunier[1]), Jacques Chatron[1]), Blandine Abs[1]), Emmanuel Ponsot[2]), Patrick Susini[3])

[1]) Aix Marseille Univ, CNRS, Centrale Marseille, LMA, Marseille, France.
  meunier@lma.cnrs-mrs.fr
[2]) Laboratoire des Systèmes Perceptifs (CNRS UMR 8248) and Département d'études cognitives,
  Ecole Normale Supérieure, PSL Research University, Paris, France
[3]) STMS (UMR9912), Ministère de la Culture, Ircam, CNRS, Sorbonne Université, Paris, France

**Summary**

The global loudness of a varying intensity sound is greater when the intensity increases than when it decreases. This global loudness asymmetry was found to be larger for pure tones than for broadband noises. In this study, our aim was to determine whether this difference between pure tones and noises is due to the difference in bandwidth between sounds or to the difference in the strength of the sensation of pitch. The loudness asymmetry was measured for broadband and for narrow-band signals that do or do not elicit a sensation of pitch. The asymmetry was greater for sounds that elicit a sensation of pitch whatever their bandwidth. The loudness model for time varying sounds [1] predicted well the asymmetry for the broadband noise that does not elicit a sensation of pitch and for a multi-tonal sound. For the other sounds the asymmetry was greater than predicted. It is known that loudness and pitch interact. The difference in asymmetry between sounds that elicit pitch and sounds that do not elicit pitch might be due to this interaction.

## 1. Introduction

Perceptual differences have been observed between sounds that increase and sounds that decrease in intensity with identical energies and long-term spectra. Rising-intensity sounds are perceived longer [2, 3] and as changing more in loudness than falling-intensity sounds [4, 5]. Their global loudness (that is, the overall loudness of the sound over its entire duration) is also greater than the global loudness of a falling sound. At the point of subjective equality (PSE), a falling 1-kHz pure tone varying over 15-dB has a rms level that is 4 dB higher than that of its symmetrical rising version [6]. This global loudness difference is usually called loudness asymmetry. Various studies have attempted to explain and model the global loudness asymmetry [7, 6]. Ponsot *et al.* [7] showed a larger loudness asymmetry for pure tones than for broadband noises. Neuhoff [5] found the same tendency when addressing loudness change. Ponsot *et al.* [8] showed that the asymmetry in loudness was not due to a different weighting of the loudest part of the signal when presented at the end rather that at the beginning of the signal.

The aim of the present study was to determine whether the difference in loudness asymmetry between tones and noises is due to their difference in spectrum width or in pitch strength. Indeed, pure tones elicit strong perception of pitch whereas broadband noises do not elicit any pitch, which may explain the difference in asymmetry. It is known that the dimensions of loudness and pitch interact with each other (see [9] for example). For example, when listeners have to perform an intensity discrimination task, the reaction time is faster when both stimuli have the same pitch or when pitch and loudness are congruent (high pitch/high loudness, low pitch/low loudness) than when pitch and loudness are incongruent (high pitch/low loudness, low pitch/high loudness). Moreover, the judged pitch of rising-intensity sound (with constant frequency) increases and the judged pitch of falling-intensity sound decreases ([10, 11]). Thus combining loudness variation (due to intensity variation) and pitch variation (also due to intensity variation) may reinforce the difference between rising and falling ramps.

In this study, we attempted to measure the loudness asymmetry for broadband and narrow-band sounds that do and do not elicit pitch, in order to separate the effects of signal bandwidth from those of pitch.

## 2. Method

Seven women and nine men took part in the experiment. All listeners had hearing thresholds less than 15 dB HL. Their thresholds were determined using standard pure-tone audiometry in the frequency range between 0.125 and 8 kHz (with an AC40 audiometer). The participants were all naïve with respect to the hypotheses under test. They were paid for their participation.

The sounds were played via a RME UCX Fireface sound-card and were presented diotically through headphones (Sennheiser HDA 200). They were sampled at a frequency of 44.1 kHz with 16 bits resolution. The headphones were calibrated using a Brüel and Kjær Artificial Ear (type 4153) coupled with the mounting plate provided for circumaural headphones with no free-field equalization and a Brüel and Kjær voltmeter (26209). The experiment took place in a double-walled soundproof booth.

The level of the stimuli varied linearly over its duration between 50 and 65 dB SPL, either increasing in intensity for the rising-intensity ramps or decreasing for the falling-intensity ramps. Four different spectral contents were created in order to obtain a broadband sound with pitch, a broadband sound without pitch, a narrow-band sound (less than 1 critical band) with pitch and a narrow-band sound without a constant pitch. The broadband sound with pitch was a frozen white noise repeated each millisecond, which created a pitch corresponding to a frequency of 1 kHz (RWN). Its spectrum is that of a harmonic complex tone with a fondamental frequency of 1 kHz, random phases and amplitudes (see [12]). The broadband sound without pitch was a white noise (20 Hz–20 kHz, WN). The narrow-band sound with pitch was a pure tone at 1 kHz (PT). Creating a narrow-band sound that does not elicit any pitch is not straightforward. Noise of bandwidth less than one critical band cannot be considered as providing no pitch. The strength of the pitch is weaker than that of a pure tone, but it exists and corresponds to the central frequency of the noise. Instead of looking for narrow-band noise without pitch, we designed a signal with multiple pitches. The idea was that listeners could not concentrate on each pitch and thus would not perceive a global pitch at the end of the sound. The stimulus was a pure tone whose frequency varied over variable periods of time. The frequency was randomly drawn on an uniform distribution (on the hertz scale) between 920 and 1080 Hz (width of one critical band). Segment durations were chosen randomly between 50 and 150 ms (Figure 1). Each segment was gated on and off (r/f time=2 ms), preventing spectral splatter. The segments were not overlapping. Once the frequency and the duration of each segment selected, the multi-tonal (MT) sound was frozen throughout the experiment. Affirming that MT does not elicit any pitch would be incorrect. Its pitch changes very quickly, it can only be said that it does not elicit a global pitch. Thus its pitch strength might be weaker than that of PT or RWN. The first three sounds lasted 2s. The multi-tonal stimulus was slightly shorter (1930.5 ms) because of the variable duration segments. The rise/fall times of all signals were 10 ms.

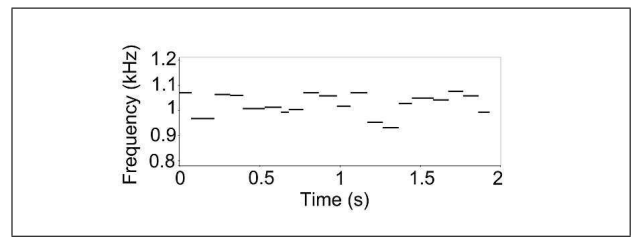

Figure 1. time-frequency representation of the multi-tonal stimulus (MT).

The loudness asymmetries were evaluated by measuring the PSE in loudness between signals with rising and falling ramps. This procedure has already been employed in previous studies ([6], [7]) and showed equivalent results to an absolute magnitude estimation task. A two-interval, two-alternative forced-choice paradigm (2I-2AFC) based on an interleaved adaptive procedure ([13], [14]) was used to measure the PSEs. On each trial, the listeners heard two sounds, one with falling and one with rising intensity, separated by 500 ms. Their task was to indicate which sound was louder by pressing a key. The response initiated the next trial after a 1-s delay. The overall level of one sound was fixed (test sound) all along the track while the overall level of the other varied (comparison sound) depending on the response of the listener to the preceding trial. The minimum and maximum levels of the comparison sound varied, while its dynamic was kept constant at 15 dB, and were adjusted according to a simple up-down procedure [15]. If the listener indicated that the comparison sound was the louder one, its maximum level (and thus its minimum level) was reduced, otherwise it was increased. At the beginning of a track the maximum (and minimum) level of the comparison sound was 5 dB higher or lower than the maximum (and minimum) level of the test sound. Step-size variation was 6 dB until the second reversal and was reduced by a ratio of 2 after every two reversals and was held constant after six reversals. A track ended when 8 reversals were achieved. The level corresponding to the matching loudness was defined as the average of the maximum level of the comparison sound in the last two reversals of each track. This procedure converges at the level corresponding to the 50% point on the psychometric function [15].

One block was composed of four interleaved tracks (Figure 2). In one track, the pair order was rising/falling and the second sound of the pair was the comparison stimulus (Figure 2a), in another track the comparison stimulus was presented first (Figure 2b). In two other tracks the pair order was falling/rising and the comparison stimulus was either presented first (Figure 2d) or second (Figure 2c). Each type of stimulus (RWN, PT, WN and MT) was tested in different blocks. Block order was randomized across listeners. Each listener completed two blocks. The asymme-try was calculated as the difference between the maximum level of the falling sound and the maximum level of the rising sound at PSE (mean of the two values obtained in each block). The asymmetry was calculated for each pair order (rising/falling and falling/rising) as the average of the asymmetry obtained for the two tracks that differed by the position of the comparison stimulus (1st or 2nd interval). Rising first and falling first are shown separately, as previous studies have shown an effect of the order of presentation on the asymmetry ([3], [7]).

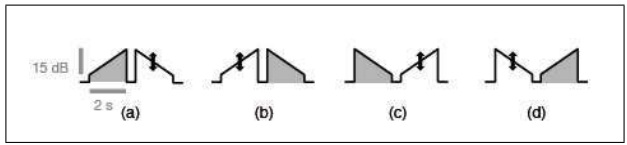

Figure 2. Experimental configuration.

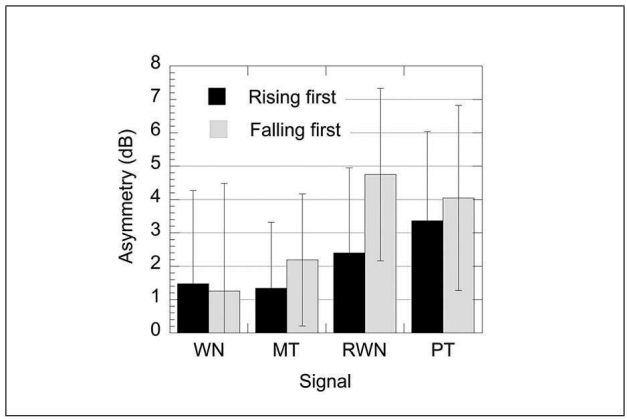

Figure 3. Loudness asymmetries between rising and falling ramps. Positive asymmetry indicates that the falling sound was matched higher in level than the rising sound in order to be perceived equally loud, which means that the falling sound would be perceived softer than the rising sound of the same level. Error bars correspond to standard deviation.

## 3. Results

Figure 3 shows the loudness asymmetries obtained with the different types of sounds and for the different pair orders. Repeated-measures ANOVAs showed a significant effect of the type of stimulus [$F(3, 15) = 15.21$, $p < 0.001$, $\eta_p^2 = 0.5$] and no effect of the pair order [$F(1, 15) = 3.02$, $p = 0.1$]. The analysis showed a significant type-of-stimulus x pair-order interaction [$F(3, 45) = 6.35$, $p = 0.001$, $\eta_p^2 = 0.3$]. Post-hoc LSD tests showed no significant difference between WN and MT ($p = 0.37$) and between PT and RWN ($p = 0.76$). The asymmetry produced by WN was significantly smaller than those produced by RWN and PT ($p < 0.001$). The asymmetry produced by MT was significantly smaller than those produced by RWN and PT ($p < 0.001$). The results show that the asymmetry is greater for the sounds producing a constant or global pitch (RWN and PT), whatever their frequency widths, than for the white noise.

## 4. Discussion

In the experiment reported here, sounds that elicit a constant or global pitch sensation produce larger loudness asymmetries than WN and MT, irrespective of bandwidth. It thus seems that pitch had an effect on the size of the loudness asymmetry, while bandwidth did not. To summarize the data, the asymmetries were averaged across pair order. The loudness asymmetry induced by WN (asymmetry=1.36 dB) was smaller than that induced by MT (asymmetry=1.76 dB) which was smaller than that induced by RWN (asymmetry=3.56 dB) which was smaller than that induced by PT (asymmetry=3.68 dB).

In order to explain the asymmetries and the differences between RWN and PT on the one hand, and WN and MT on the other hand, the loudness model for time varying sounds of Glasberg and Moore [1] was used to predict loudness differences between the rising- and falling-intensity sounds used in our study. The difference in level between rising and falling sounds needed to obtain the same peak value of the Long Term Loudness (LTL) for both signals was used to estimate the asymmetry in dB. The predicted asymmetry was 1.3 dB for WN, 2.1 dB for MT, 1.5 dB for RWN and PT. The observed asymmetry was quite well explained by the model for the sound that does not elicit pitch (WN) whereas it was underestimated for the sounds that elicit pitch (RWN and PT). For MT, the loudness model slightly overestimated the asymmetry. Our results are in agreement with those of Ries *et al.* [3] who found that LTL predicted well the loudness asymmetry for white noise of 500 ms, and with those of Ponsot *et al.* [6], who predicted an asymmetry of 1.24 dB for a 1-kHz pure tone of 2 s varying between 55 and 70 dB, using the LTL model.

The long-term loudness is calculated from instantaneous loudness using two stages of temporal integration. It models the overall loudness impression which is probably formed at a central level of the auditory system. For the white noise, it seems that this model based on temporal integration is consistent with the behavioral data. However, for the repeated white noise and the 1-kHz pure tone the model underestimated the loudness asymmetry. Our assumption is that the loudness asymmetry is increased by the sensation of pitch. McBeath and Neuhoff [11] showed that tones with continuous intensity change and constant frequency were perceived as changing in pitch. When the intensity increased (respectively decreased), the pitch increased (respectively decreased). Moreover, the pitch change of rising-intensity sounds was larger than the pitch change of falling-intensity sounds. The combined effect of loudness variation (due to intensity variation) and pitch variation (also due to intensity variation) might reinforce the loudness difference between sounds with rising and falling ramps. Thus, for RWN and PT the asymmetry would result from temporal integration of instantaneous loudness and a cognitive mechanism due to the interaction between the dimensions of loudness and pitch. For MT, we could emit the same hypothesis as for WN, but its peculiarity (several pitches) leads us to remain cautious. It should also be noted that the difference between WN and RWN might be due to the fact that WN has more energy in the low frequency region than RWN. The asymmetries were assessed with the same participants for the different sounds. This allows calculation of interindividual correlations between these conditions; significant correlations between conditions might indicate shared mechanisms [16, 17]). Non-parametric correlations between the 6 possible combinations (Kendall's tau) were all significant at $p < 0.05$ (after Bonferonni correction with alpha = 6), except those with the MT condition. Thus subjects with greater asymmetries, e.g. for PT, also exhibited stronger asymmetries for WN and RWN. This might indicate that the mechanism(s) at the origin of the asymmetries for the white noise, the repeated white noise as well as the 1-kHz pure tone is (are) shared (at least partially), whereas other mechanism(s) might contribute to those observed for the multi-tonal sound, reinforcing its peculiarity. These assumptions could be tested in further studies in which the loudness asymmetry would be measured for sounds with parameterizable pitch strength (like iterated rippled noises or pure tone in noise) and the correlation between pitch strength and loudness asymmetry would be assessed.

## References

[1] B. R. Glasberg, B. C. J. Moore: A model of loudness applicable to time-varying sounds. J. Audio Eng. Soc. **50**(5) (2002) 331–342.

[2] M. Grassi, C. J. Darwin: The subjective duration of ramped and damped sounds. Perception and Psychophysics **68**(8) (2006) 1382–1392.

[3] D. T. Ries, R. S. Schlauch, J. J. DiGiovanni: The role of temporal-masking patterns in the determination of subjective duration and loudness for ramped and damped sounds. J. Acoust. Soc. Am. **124**(6) (2008) 3772–3783.

[4] K. N. Olsen, C. J. Stevens, J. Tardieu: Loudness Change in Response to Dynamic Acoustic Intensity. J. Exp. Psychol.-Hum. Percept. Perform. **36**(6) (2010) 1631–1644.

[5] J. G. Neuhoff: Perceptual bias for rising tones. Nature **395**(6698) (1998) 123–124.

[6] E. Ponsot, P. Susini, S. Meunier: A robust asymmetry in loudness between rising- and falling-intensity tones. Attention, Perception, & Psychophysics **77**(3) (2015) 907–920.

[7] E. Ponsot, S. Meunier, A. Kacem, J. Chatron, P. Susini: Are Rising Sounds Always Louder? Influences of Spectral Structure and Intensity-Region on Loudness Sensitivity to Intensity-Change Direction. Acta Acustica united with Acustica **101**(6) (2015) 1083–1093.

[8] E. Ponsot, P. Susini, G. Saint Pierre, S. Meunier: Temporal loudness weights for sounds with increasing and decreasing intensity profiles. J. Acoust. Soc. Am. **134**(4) (2013) EL321–EL326.

[9] R. D. Melara, L. E. Marks: Perceptual primacy of dimensions: Support for a model of dimensional interaction. Journal of Experimental Psychology **16**(2) (1990) 398–414.

[10] J. G. Neuhoff, M. K. McBeath: The Doppler illusion: The influence of dynamic intensity change on perceived pitch. Journal of Experimental Psychology: Human Perception and Performance **22**(4) (1996) 970–985.

[11] M. K. McBeath, J. G. Neuhoff: The Doppler effect is not what you think it is: Dramatic pitch change due to dynamic intensity change. Psychon Bull Rev **9**(2) (2002) 306–313.

[12] R. M. Warren, J. A. Bashford: Production of white tone from white noise and voiced speech from whisper. Bull. Psychon. Soc. **11**(5) (1978) 327–329.

[13] W. Jesteadt: An adaptive procedure for subjective judgments. Perception & Psychophysics **28**(1) (1980) 85–88.

[14] M. Florentine, S. Buus, T. Poulsen: Temporal integration of loudness as a function of level. J. Acoust. Soc. Am. **99**(3) (1996) 1633–1644.

[15] H. Levitt: Transformed Up/Down Methods in Psychoacoustics. J. Acoust. Soc. Am. **49** (1971) 467–477.

[16] G. Canévet, B. Scharf, M.-C. Botte: Simple and Induced Loudness Adaptation. Audiology **24**(6) (1985) 430–436.

[17] D. Pressnitzer, J.-M. Hupé: Temporal Dynamics of Auditory and Visual Bistability Reveal Common Principles of Perceptual Organization. Current Biology **16**(13) (2006) 1351–1357.